# LEVEL OF PRESENCE IN TEAM-BUILDING ACTIVITIES: GAMING COMPONENT IN VIRTUAL ENVIRONMENTS

Gianluca De Leo[1,2], Koren S. Goodman[1], Elena Radici[2], Scott R. Secrhist[2], Thomas W. Mastaglio[3]

[1]Virginia Modeling Analysis and Simulation Center, Old Dominion University
Norfolk, Virginia USA
`gdeleo@odu.edu,kgood006@odu.edu`

[2]School of Medical Laboratory and Radiation Sciences, Old Dominion University
Norfolk, Virginia USA
`ssechris@odu.edu, eradici@odu.edu`

[3]MYMIC LLC, Portsmouth Virginia USA
`tom.mastaglio@mymic.net`

*ABSTRACT*

*Historically the training of teams has been implemented using a face-to-face approach. In the past decade, on-line multiuser virtual environments have offered a solution for training teams whose members are geographically dispersed. In order to develop on effective team training activity, a high sense of presence among the participant needs to be reached. Previous research studies reported being able to reach a high level of presence even when using inexpensive technology such as laptop and headset. This study evaluates the level of presence of ten subjects who have to perform a team-building activity in a multi-user virtual environment using a laptop computer and a headset. The authors are interested in determining which user characterizes, such as gender, age and knowledge of computers, have a strong correlation with the level of sense of presence. The results of this study showed that female participants were more likely to engage in the activity and perceived fewer negative effects. Participants who reported less negative effects such as feeling tired, dizzy, or experiencing eye strain during the team-building activity reached a higher level of sense of presence.*

*KEYWORDS*

*Serious games, level of presence, team-building, virtual enviroments*

## 1. INTRODUCTION

Sense of presence is the subjective sensation of "being there" that users experience when they emotionally and intellectually engage in an interactive virtual environment [1][2]. Because sense of presence is the cornerstone of virtual reality (VR), understanding the level of realism that is present affects the transfer of knowledge or skill. Level of presence is a temporary experimental state, susceptible to the individuals' level of perception in an immersive environment [1][2]. The level of sense of presence perceived by users of virtual reality is influenced by media characteristics (objective presence) and user characteristics (subjective presence) [2][3]. User characteristics include fundamental factors such as age and gender, which play a major role in determining the sense of presence experienced in virtual environments [4].

Other dimensions influencing sense of presence include spatial presence, engagement, ecological validity, and negative effects [4]. Exploring the user's experiences is a useful measure when studying level of presence [5].

Substantial resources are invested annually in team building activities to enhance performance [6]. The main goal is to create a greater sense of teamwork while improving decision-making abilities among team members. However, team members may not always share the same training environment due to the increase in costs related to travel as well as the international distribution of organizations. Limitations can be overcome by using a virtual environment in which team members can meet in a virtual world and benefit from training in a shared environment. Such teams are identified as distributed teams [7]. Distributed teams are teams geographically dispersed in which they do not have direct face-to-face contact with each other [7] [8]. An example of a distributed team is the Expeditionary Medical Support (EMEDS) team. Composed of highly qualified physicians, nurses, medical technicians and administrative professionals, these team members are often exposed to potentially traumatic events [9][10]. Coordinated responses and efficient communication are vital components needed to successfully accomplish the mission of the operation. What remains a challenge for teams such as EMEDS, is the training needed to gain confidence in other team members during mission operations and the ability to maintain an expected level of vigilance when in stressful situations. Participation in simulated scenarios that elicit traumatic experiences in the virtual environment facilitates realistic team performance. A study exploring immersion in the sense of presence found that users who wore a head mounted display (HMD) to move within a virtual environment experienced negative effects [4].

This study uses a simple team building activity that has to be performed by teams in a virtual environment. Team members use a laptop computer and headset to access and navigate the virtual environment. The purpose of this study is two-fold: (1) to evaluate the level of presence when team building activities are implemented in a multi-user virtual environment with a game component using limited input devices such as a laptop computer and a headset and (2) to profile the team members to determine if there is a correlation between the users characteristics and their level of presence.

## 2. MATERIALS AND METHODS

### 2.1. Participants

Participants were recruited among the student body of Old Dominion University in Norfolk, Virginia. Inclusion criteria were: being 18 years of age or older and having the ability to use a keyboard and a mouse. The participants were randomly assigned to two teams, Virtual Group 1 and Virtual Group 2, which consisted of five members each. There were three females and two males on each team. The participants did not know each other prior to the study. Participation was voluntary, and those participating did not receive any compensation.

### 2.2. Measures

#### 2.2.1. ITC-Sense of Presence Inventory (ITC-SOPI)

The ITC-SOPI questionnaire is a post-test self-report questionnaire used to measure the sense of presence during and after a virtual experience. This assessment is composed of 44 items divided in two parts [5]. Part A has six items that measure the user impressions or feelings upon completion of the media experience, while Part B uses 38 items to measure user impressions or feelings during the media experience. The questionnaire measures four dimensions: Physical Space or Spatial Presence (19 items), Engagement (13 items), Ecological Validity (5 items), and Negative Effects (6 items). Spatial presence refers to the physical environment [5]. The content from this dimension emphasizes the level of interaction the participant experienced within the displayed environment. Engagement is the participant's intellectual involvement [5]. Participants respond to a series of statements regarding the level of involvement and whether they "liked or enjoyed" the experience. A response regarding whether the participant "felt that the displayed environment was part of the real world" is a sample content statement from the Ecological Validity dimension. Ecological Validity refers to the participant's perception of the environment [5]. Each of the four dimensions is likely to be determined by the interaction between media form, media content, and user characteristic variables. These 44 items use a 1 to 5 point Likert scale (1 corresponds to "strongly agree" and 5 corresponds to "strongly disagree").

### 2.2.2. Galvanic Skin Response

A galvanic skin response (GSR) device was used to monitor and collect the level of stress of two participants by measuring changes in the conductance of their skin pores at baseline and throughout the team building activity [11]. GSR devices have been used in the past to record physiological data [12]. A Calmlink Biofeedback GSR2 personal biofeedback device was used in this study.

### 2.3. GaMeTT Virtual Environment

Games for Team Training (GaMeTT), is a multi-player, online, virtual environment used for team training [13]. GaMeTT is built on Olive, a commercial software able to develop and deliver an accessible virtual world using standard computing technology via the World Wide Web [14]. Users are not required to have expert computer/technology skills to use GaMeTT. GaMeTT allows each individual the opportunity to interact within the virtual environment along with fellow team members via a realistic avatar representation of their physical appearance [13]. Avatar movements are controlled using the mouse and the arrow buttons on the keyboards. Users have the capability to communicate using a headset and a microphone.

### 2.4. Hardware

GaMeTT was installed on five laptop computers with a 15-inch monitor. The virtual environment was viewed on each laptop in a full screen mode. Each avatar had the name of the participant above its head (See Figure 1). Avatar movements in the virtual environment were controlled using the arrows on the keyboard, while the mouse allowed the user to change the perspective of the virtual environment.

### 2.5. Procedure

The team building activity required each team member to use the mouse and arrows to step onto 30 numbered markers randomly placed in a virtual hotel courtyard in sequential order from 1-30 as quickly as possible. When the avatars moved onto the marker, the participants had to shout out the number on which they stepped. This allowed the other team members the chance to move towards the next number. When an avatar stepped onto a marker in the wrong sequence, an error was counted and the team had to restart the task from the beginning. Both teams were allowed 10 minutes to complete the task. This timeframe allowed team members to plan and devise strategies as often as necessary for improved performance. Each virtual experience performance was recorded to capture the number of attempts and the planning time of each team for analysis. Logistics and space available for this research study accommodated five members on each team. Prior to the beginning of the research study, in order to allow participants to practice how to navigate inside the virtual environment, a 10 minutes training session was provided. Participants were free to talk each other and to navigate in a virtual environment representing a conference room. The participants were part of a larger research project that found a high transfer of knowledge and skill training among distributed teams within virtual training environments by measuring cohesiveness and team performance [15].

This research received approval from the Human Subjects Institutional Review Board of Old Dominion University. Analyses were performed using SPSS 17.0. All p-values reported in this study are two-tailed and α<.05 is considered statistically significant.

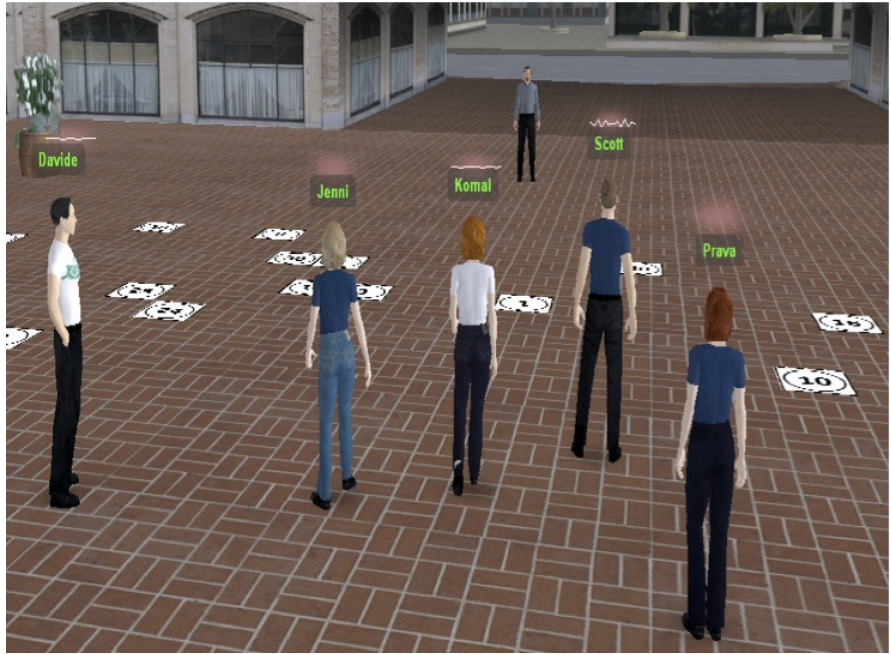

Figure 1. Virtual Environment View

## 3. RESULTS

The participants (n=10) in the study were predominately female (60% female; 40% male). The mean age of the sample was 27.50 (SD=8.8). More than half of the participants were less than 25 years old (n=6) and reported having earned a college degree. Of those participating, 60% had a basic level of knowledge of virtual reality, 50% affirmed having a basic level of knowledge of how 3-D images are produced, and 80% previously viewed stereoscopic 3-D images using polarized glasses. Of the two teams, members of Virtual Group 1 were slightly older (M=31.2, SD=11.3) than Virtual Group 2 (M= 23.8, SD=3.4). Participants reported having an intermediate level of computer experience. Both groups had similar experiences with virtual reality systems, and the majority (n=9) reported having never used a virtual reality system before the research study. (See Table 1)

Table 1. Participant Demographics

| Participant Demographics | N | % |
|---|---|---|
| **Age** | | |
|   20-25 | 6 | 60% |
|   26-30 | 2 | 20% |
|   31-35 | 1 | 10% |
|   >35 | 1 | 10% |
| **How often you play computer games** | | |
|   Never | 1 | 10% |
|   Occasionally | 5 | 50% |
|   Often but less than 50% of days | 3 | 30% |
|   50% or more of days | 1 | 10% |
| **Average weekly TV viewing** | | |
|   0-8 hours | 8 | 80% |
|   9-16 hours | 2 | 20% |
| **Education** | | |
|   HS Diploma | 3 | 30% |
|   College Degree | 7 | 70% |
| **Level of knowledge of virtual reality** | | |
|   None | 4 | 40% |
|   Basic | 6 | 60% |
| **Viewed 3-D stereoscopic images using polarized gasses** | | |
|   Yes | 8 | 80% |
|   No | 2 | 20% |
| **Used experimental virtual reality before** | | |
|   Yes | 9 | 90% |
|   No | 1 | 10% |
| **Level of knowledge of how 3-D images are produced** | | |
|   None | 4 | 40% |
|   Basic | 5 | 50% |
|   Intermediate | 1 | 10% |

The analysis of variance (ANOVA) revealed the two groups reported similar responses when asked about the average number of hours spent weekly on TV viewing, the level of knowledge of TV/Film production, and whether the participant previously viewed 3-D stereoscopic images using polarized glasses. The two groups were composed by the same number of males and females. The two groups did not differ significantly on other user characteristics as shown in Table 2.

The International Journal of Multimedia & Its Applications (IJMA) Vol.3, No.2, May 2011Table 2. ANOVA for Background Questionnaire for Virtual Group 1 and Virtual Group 2

| User Characteristics for Virtual Group 1 and Virtual Group 2 Team | p-value |
|---|---|
| Age | .197 |
| Education | .545 |
| Gender | 1.000 |
| Average Weekly TV Viewing | 1.000 |
| Viewed 3-D stereoscopic images using polarized gasses | 1.000 |
| Level of knowledge of TV/Film Production | 1.000 |
| TV Size | .667 |
| Time spent playing with a video | .486 |
| Used experimental virtual reality before | .347 |
| Computer Experience | .347 |
| Level of knowledge of Virtual Reality | .242 |
| Level of knowledge of how 3-D images are produced | .172 |

The results of the ITC-SOPI Questionnaire for each team member were combined to develop a mean score for each team. The means and standard deviations of the four scales of the ITC-SOPI questionnaire for each team revealed the Spatial Presence scale had the same mean (3.2) for the two groups. When measuring Engagement and Ecological Validity, Virtual Group 2 revealed higher means (Engagement, 3.7; Ecological Validity, 3.3) compared to those found in Virtual Group 1(Engagement, 3.3; Ecological Validity, 3.2) (See Table 3). When measuring the means of Negative Effects between the two groups, Virtual Group 2 had a lower mean score (1.2) compared to Virtual Group 1 (2.1). However, the results of the ANOVA confirmed there were no significant differences among the two groups when measuring the four scales of the ITSC-SOPI (Spatial Presence, p=.979; Engagement, p=.601; Ecological Validity, p=.822; Negative Effects, p=.226). Table 3 details the results of the sense of presence for individual team performance and the combined means and standard deviations.

Table 3. Virtual Team Mean Values and Standard Deviations of ITC-SOPI Questionnaire

| Team | Spatial Presence | | Engagement | | Ecological Validity | | Negative Effects | |
|---|---|---|---|---|---|---|---|---|
| | Mean | SD | Mean | SD | Mean | SD | Mean | SD |
| **Virtual Group 1** | 3.2 | 1.3 | 3.3 | 1.1 | 3.2 | 0.9 | 2.1 | 1.4 |
| **Virtual Group 2** | 3.2 | 1.0 | 3.7 | 0.8 | 3.3 | 1.2 | 1.2 | 0.2 |
| **Combined** | 3.3 | 1.1 | 3.5 | 0.9 | 3.2 | 1.0 | 1.7 | 1.1 |

There were significant differences among gender when measuring for Spatial Presence and Ecological Validity. Female participants were more likely to feel immersed in the environment compared to males (Females, M=3.6, SD=0.8 v. Males, M=2.6, SD=1.2) (See Table 4). The results also revealed that female participants believed the environment to be natural (M=3.8, SD=0.9 v. M=2.5, SD=0.4). Females report that the team building exercise could be easily executed in the real world. However, males were less likely to engage compared to their female counterparts (Males, M=2.8, SD=0.8 v. Females, M=4.0, SD=0.6). Males also experienced

The International Journal of Multimedia & Its Applications (IJMA) Vol.3, No.2, May 2011more negative effects (M=1.9, SD=1.8). Overall, the results revealed the participants responded with emotion and felt a connection with other avatars in the virtual world. Significant differences were observed on the Engagement scale and in the Ecological Validity scale among both female and male participants. The p-values were .029 and .041 respectively. Spatial Presence (p=.138) and Negative Effects (p=.566) were not significant.

The results did not reveal any statistically significant differences among age and the four scales. Higher mean scores were reported among those participants less than 25 when measuring Spatial Presence, Engagement, and Ecological Validity (See Table 4). Participants 26 years and older had higher mean scores when asked if they felt disorientated after the experience compared to those less than 25.

ANOVA tests were conducted between each of the user characteristics information and the ITC-SOPI four scales. See Table 5. The only participant who reported an expert level of TV/Film production knowledge experienced more negative effects, while those who self-report having a basic level of TV/Film production knowledge experienced minimal negative effects. Significant differences were found when measuring computer experience (p=.043), TV and film production knowledge (p < .001), and the level of knowledge of how 3-D images are produced (p=.009). Participants with a basic level of computer experience felt the content depicted a real world environment.

Table 4. Mean Values Calculated for Each Variable

|  | Spatial Presence | Engagement | Ecological Validity | Negative Effect |
|---|---|---|---|---|
| Gender | | | | |
| **Female** | 3.6 ± 0.8 | 4.0 ± 0.6 | 3.8 ± 0.9 | 1.5 ± 0.4 |
| **Male** | 2.6 ± 1.2 | 2.8 ± 0.8 | 2.5 ± 0.4 | 1.9 ± 1.8 |
| Age | | | | |
| **Less than 25** | 3.6 ± 0.9 | 3.7 ± 0.8 | 3.3 ± 1.2 | 1.3 ± 0.3 |
| **26 and older** | 2.6 ± 1.3 | 3.1 ± 1.1 | 3.1 ±0.9 | 2.1 ± 1.7 |

The galvanic skin response device was used to measure the stress level of two participants, one member from each team. Sensors were applied to the left hand, using the first two fingertips to record changes in skin conductance [11]. The galvanic skin response device needs to be calibrated every time it is used on a different subject. Since the calibration is different the results obtained from the first participant cannot be compared to the second participant. The results are meaningful only if compared between baseline and experiment. Data collected from the two participants occurred at two separate intervals, at baseline, five to seven minutes prior to the start of the study and during the activity that lasted approximately 12 minutes.

Table 5.  ANOVA Results of Background Questionnaire of ITC-SOPI

| Items Background Questionnaire | Items | N | Mean per item | | ANOVA Results |
|---|---|---|---|---|---|
| Rate your level of computer experience | None<br>Basic<br>Intermediate<br>Expert | 0<br>3<br>6<br>1 | Ecological Validity<br><br>4.2 ± 0.72<br>2.6 ± 0.75<br>4.2 | | Ecological Validity<br>F=5.073<br>p=.043 |
| How would you rate your level of TV/film production knowledge | None<br>Basic<br>Intermediate<br>Expert | 0<br>6<br>3<br>1 | Negative Effects<br><br>1.5 ± 0.46<br>1.1 ± 0.09<br>4.7 | | Negative Effects:<br>F=34.734<br>p= < .001 |
| How would you rate your knowledge of how 3D images are produced | None<br>Basic<br>Intermediate<br>Expert | 4<br>5<br>1<br>0 | Engagement<br><br>4.1 ± 0.50<br>3.2 ± 0.58<br>1.7 | Negative Effects<br><br>1.4 ± 0.39<br>1.3 ± 0.48<br>4.7 | Engagement:<br>F= 9.956<br>p= .009<br><br>Negative Effects:<br>F=25.742<br>p= < .001 |

Mean values were recorded and revealed that one participant in Virtual Group 1 showed a significant difference in skin conductance between baseline measures and during the team building exercise (M=3978, SD=1779 vs. M=7376, SD=3298). Specifically for this participant, when asked to begin the team building exercise, the results show an immediate increase in the stress level.  This increase remained significantly higher for the length of the team building exercise compared to that of the other participant.  A slight decrease was noted in the mean stress level value recorded for the participant in Virtual Group 2 at baseline and during the team building exercise (M=8879, SD=3971 vs. M=8708, SD=3894).

## 4. CONCLUSION

The literature on the sense of presence places emphasis on studying the characteristics of the media used and the content of the virtual environment with the goal of determining how much the users feel immersed in a virtual environment. Previous research supports considering user characteristics as a vital component when measuring the level of presence [16]. What remains a challenge is to determine which user characteristics make a virtual team training more effective. In order to facilitate and to maximize the learning outcome of a team such as an EMEDS, it would be helpful to select participants for whom the virtual training is known to be more effective and beneficial.

In this study, female participants were more likely to engage in the activity and perceived fewer negative effects. A study exploring virtual team perceptions and user interactions and experiences confirmed that women experience a greater level of satisfaction while working in virtual groups compared to men [17]. Previous findings also suggest that women perceived that virtual groups displayed a greater level of camaraderie compared to men [17]. While the ANOVA results of this research study did not reveal a statistically significant difference between the age and the sense of presence, those less than 25 were more susceptible to the sense of presence compared to those 26 years and older. A previous study did not find any correlation between sense of presence and age [17].

Of the participating teams, Virtual Group 2 members manifested a higher level of sense presence during the execution of the virtual activity compared to Virtual Group 1. Virtual Group 2 members were also more likely to engage in the team building activity and considered the virtual environment more natural and realistic compared to members of Virtual Group 1. The Negative Effects scale may explain the differences between the groups, as Virtual Group 1 were more likely to feel tired, dizzy, or experience eye strain, nausea, and headaches during the experience. A previous study suggests these symptoms are inversely related [18].

A study exploring the role of sense of presence on cognitive rehabilitation report that changes in skin conductance of those participating in an immersive virtual environment may be the result of increased levels of intellectual involvement in the task and the perceived level of difficulty of the task [11]. Although this study found that one participant experienced higher levels of stress between baseline measures and during the team building exercise, future studies should focus on the relationship between user characteristics and stress levels to determine the factors that predict higher levels of stress. Specifically for distributed teams such as EMEDS, which are composed of multidisciplinary members, performing in an immersive virtual environment will allow the user to identify specific events that elevate stress so that a level of vigilance is maintained in a real life scenario.

Although previous research has demonstrated significant findings on sense of presence using similar sample sizes [19], the major limitation of this study is the small sample size. Another limitation of this is that the demographic of the sample is not representative of a larger population and, therefore, limits the generalization of results.

Previous research studies revealed that because helmets and sensors produce some negative effects related to motion, technologies such as a laptop computer and a headset support a good level of sense of presence and are capable of being used in a distributed team training environment [4]. The present study evaluated the level of presence when team-building activities are implemented in a multi-user virtual environment using laptops and headsets. The findings of this study suggest that for distributed teams such as EMEDS, placing emphasis on user characteristics will provide a foundation to create mission oriented training modules while accounting for individual demographic variables such as gender and age differences and the familiarity with computer technology or usage of TV. In order to study more in detail the user characteristics, future research studies should use EMEDS personnel and give them the opportunity to train in a virtual training session that would be very similar in terms of virtual environments and tasks to the real EMEDS operations.

## ACKNOWLEDGEMENTS

This research was sponsored by the US Air Force Research Laboratory.

## REFERENCES

[1] Usoh, M., Ernest, C., Arman, S., & Slater, M. (2000). Using presence questionnaires in reality. *Presence: Teleoperators and Virtual Environments*, 9(5), 497-503.

[2] Slater, M., Pertaub, D., & Steed, A. (1999). Public speaking in virtual reality: facing an audience of avatars. *IEEE Computer Graphics and Applications*, 19(2), 6-9

[3] Slater, M. & Wilbur, S. (1997). A framework for immersive virtual environments (FIVE): speculations on the role of presence in virtual environments. *Presence: Teleoperators and Virtual Environments*, 6(6), 603-16.

[4] Banos, R.M., Botella, C., Alcaniz, M., Liano, V., Guerrero, B., & Rey, B. (2004). Immersion and emotion: their impact on the sense of presence. CyberPsychology & Behavior , 7(6), 734-40.

[5] Lessiter, J., Freeman, J., Keogh, E., & Davidoff, J. (2001). A cross-media presence questionnaire: the ITC-sense of presence inventory. *Presence: Teleoperators and Virtual Environments*, 10(3), 282-97.

[6] Bohl, D.L., Slocum, J.W., Luthans, F., & Hodgetts, R.M. (1996). Ideas that will shape the future of management practice. *Organizational Dynamics*, 25(1), 7-14.

[7] Fiore, S.M., Salas, E., Cuevas, H.M., & Bowers, C.A. (2003). Distributed coordination space: toward a theory of distributed team process and performance. *Theoretical Issues in Ergonomics Science*, 4(3-4), 340-64.

[8] Hammond, J.M., Harvey, C.M., Koubek, R.J., Compton, W.D., & Darisipudi, A. (2005). Distributed collaborative design teams: media effects on design processes. International. *Journal of Human Computer Interaction* 2005, 18(2), 145-65.

[9] Collins, S. (2008).Emergency medical support units to critical care transport teams in Iraq. *Critical Care Nursing Clinics of North America*, 20(1), 1-11.

[10] Skelton, P.A., Droege, L., & Carlisle, M.T. (2003). EMEDs and SPEARR teams: United States Air Force ready responders. *Critical Care Nursing Clinics of North America*, 15(2), 201–12.

[11] Lo Priore, C., Castelnuovo, G., Liccione, D., & Liccione, D. (2003). Experience with v-store: considerations on presence in virtual environments for effective neuropsychological rehabilitation of executive functions. *CyberPsychology & Behavior*, 6(3), 281-87.

[12] Starcke, K., Tuschen-Caffier, B., Markowitsch, H.J., & Brand, M. (2009). Skin conductance responses during decisions in ambiguous and risky situations in obsessive-compulsive disorder. *Cognitive Neuropsychiatry*, 14(3), 199-216.

[13] MYMIC LLC (MYMIC Limited Liability Company). (2010). Retrieved online September 30, 2010 at http://www.mymic.net/media/GaMeTT%20Flyer.pdf.

[14] Forterra Systems Incorporated. (2010). Retrieved online August 25, 2010 at http://www.forterrainc.com/index.php/product-a-services.

[15] De Leo, G., Sechrist, S., Radici, E., & Mastaglio, T.W. (2010). Games for Team Training. *Interservice/ Industry Training, Simulation and Education Conference (I/ITSEC)*, November 28-December 1, 2010, Orlando, FL, 2327-2336.

[16] Steuer, J. (1992). Defining virtual reality: dimensions determining telepresence. *Journal of Communication*, 42(4),73-93.

[17] Lind, M.R. (1999). The gender impact of temporary virtual work groups. *IEEE Transactions on Professional Communication*, 42(9), 276-84.

[18] Witmer, B.G. & Singer, M.J. (1998). Measuring sense of presence in virtual environments: a presence questionnaire. *Presence: Teleoperators and Virtual Environments*, 7(3), 225-40.

[19] Slater, M., Pertaub, D., & Steed, A. (1999). Public speaking in virtual reality: facing an audience of avatars. *IEEE Computer Graphics and Applications*, 19(2), 6-9.